\documentstyle[12pt]{article}

\catcode`\@=11

\def\theequation{\arabic{equation}}

\def\@normalsize{\@setsize\normalsize{15pt}\xiipt\@xiipt
\abovedisplayskip 14pt plus3pt minus3pt%
\belowdisplayskip \abovedisplayskip
\abovedisplayshortskip  \z@ plus3pt%
\belowdisplayshortskip  7pt plus3.5pt minus0pt}

\def\small{\@setsize\small{13.6pt}\xipt\@xipt
\abovedisplayskip 13pt plus3pt minus3pt%
\belowdisplayskip \abovedisplayskip
\abovedisplayshortskip  \z@ plus3pt%
\belowdisplayshortskip  7pt plus3.5pt minus0pt
\def\@listi{\parsep 4.5pt plus 2pt minus 1pt
            \itemsep \parsep
            \topsep 9pt plus 3pt minus 3pt}}

\def\underline#1{\relax\ifmmode\@@underline#1\else
        $\@@underline{\hbox{#1}}$\relax\fi}
\@twosidetrue

\relax

\catcode`@=12

\catcode`\@=11

\def\section{\@startsection{section}{1}{\z@}{3.5ex plus 1ex minus
   .2ex}{2.3ex plus .2ex}{\large\bf}}

\def\thesection{\Roman{section}.}

\def\appendix{\setcounter{section}{0}
        \def\thesection{Appendix}
        \def\theequation{\Alph{section}.\arabic{equation}}}

\def\ps@headings{\def\@oddfoot{}\def\@evenfoot{}
\def\@oddhead{\hbox{}\hfill
        \makebox[.5\textwidth]{\raggedright\ignorespaces --\thepage{}--
        \hfill {}}}
\def\@evenhead{\@oddhead}
\def\subsectionmark##1{\markboth{##1}{}}
}

\ps@headings

\catcode`\@=12

\relax

\def\figcap{\section*{Figure Captions\markboth
        {FIGURECAPTIONS}{FIGURECAPTIONS}}\list
        {Fig. \arabic{enumi}:\hfill}{\settowidth\labelwidth{Fig. 999:}
        \leftmargin\labelwidth
        \advance\leftmargin\labelsep\usecounter{enumi}}}
 \relax
\def\tablecap{\section*{Table Captions\markboth
        {TABLECAPTIONS}{TABLECAPTIONS}}\list
        {Table \arabic{enumi}:\hfill}{\settowidth\labelwidth{Table 999:}
        \leftmargin\labelwidth
        \advance\leftmargin\labelsep\usecounter{enumi}}}
 \relax
\def\reflist{\section*{References\markboth
        {REFLIST}{REFLIST}}\list
        {[\arabic{enumi}]\hfill}{\settowidth\labelwidth{[999]}
        \leftmargin\labelwidth
        \advance\leftmargin\labelsep\usecounter{enumi}}}
 \relax

\catcode`\@=11

\def\ps@headings{\def\@oddfoot{}\def\@evenfoot{}
\def\@oddhead{\hbox{}\hfill
        \makebox[.5\textwidth]{\raggedright\ignorespaces --\thepage{}--
        \hfill {}}}
\def\@evenhead{\@oddhead}
\def\subsectionmark##1{\markboth{##1}{}}
}

\ps@headings

\relax

\newskip\humongous \humongous=0pt plus 1000pt minus 1000pt

\newif\ifdtup

\def\beq{\begin{equation}}
\def\eeq{\end{equation}}

\def\beqn{\begin{eqnarray}}
\def\eeqn{\end{eqnarray}}
\relax

\def\G2{{\; \rm GeV/}c^2}
\def\G{\; \rm GeV}

\def\dotx{\dotx{\dot\overline{x}}}

\relax

\begin{document}
\begin{titlepage}
\nopagebreak
\begin{flushright}

        {\normalsize
                        ITP--SB--91-47\\
                         September,~1991\\ }

\end{flushright}

\vfill
\begin{center}
{\large\bf
   Matrix Models
at
     Finite  $N$} \footnote{ based on the talk given at the conference
 ``Strings and Symmetries 1991'' at Stony Brook, May 20-25, 1991.}

\vfill
\vskip 0.35in
       {\bf H.~Itoyama}\footnote{work supported in part by
        NSF Grant Phy 91-08054}   \\
 Institute for Theoretical Physics,\\
        State University of New York at Stony Brook,\\
        Stony Brook, NY 11794-3840\\

\end{center}
\vfill

\begin{abstract}
  We summarize some aspects  of matrix models from the approaches directly
based on their properties  at finite $N$.

\end{abstract}
\vfill
\end{titlepage}

\section{Introduction}
\label{Introduction}

   Current interest in string theory in less than
or equal to one dimension has arisen from the discovery of the existence of the
 continuum limit called double scaling limit \cite{DSL}
  which sums   string perturbation
theory. At the same time, it has become
 clear that the universal equations which
govern the correlation functions near the critical points are given by
hierarchical integrable differential equations \cite{DSL,Douglas,Rutgers}.

 By some time last year, however,  more than several people had come to
 realize
 that many of the results  obtained in the limit are
already visible  when  $N$- the size of the matrix- is kept finite.
This provides an opportunity to study the system in its original definition
and as a solvable model. The progress from that direction  is what  we will
discuss below.
We  will summarize  some aspects
 of zero dimensional matrix models
  from the point of view based directly on  results obtained at finite $N$.

Such analyses appear to be imperative in view of the recent status
of nonperturbative $2d$ gravity: no limiting procedure has been
found, up to now,
 which maintains both the reality of the partition function and the original
combinatorial correspondence of a matrix model with a triangulated two
dimensional surface \cite{David,Marinari}.
No satisfactory definition of continuum $2d$
 gravity ( nonperturbative) has, therefore, been given.

 In the one matrix model, we will cover the following items:
 Virasoro constraints at finite $N$ fully characterize the space of
 correlators: the equivalence of the two approaches, i.e.
 Dyson-Schwinger approach
and the one based on orthogonal polynomials;  how to take the double
scaling limit of the Virasoro constraints at finite $N$ near the
 $(2,2k-1)$ critical points \cite{IM} to obtain the  Virasoro constraints
 of a twisted boson \cite{VV,TOKYO};
correspondence between the one matrix model and the classical
 Toda lattice equation \cite{Toda}:
 the
   derivation of Kazakov's loop equation \cite{Kazakov} directly
from  the Virasoro constraints  (See, for instance, \cite{AIM}).
See also \cite{MIS} for these items.

 In the case of the two matrix model,  the situation
has been clarified since the original developments and
 proposals \cite{Douglas,Ising}. By now,
we agree that the two matrix models capture the essential feature of the
2d gravity coupled to a general $(p,p^{\prime})$ conformal
  matter \cite{TY,DD}: it
contains all the
 critical points indexed by a set of coprime integers $(p,p^{\prime})$.
 We will discuss the new constraints of $w_{\infty}$
type derived in  \cite{IM2}.

\subsection{One matrix model}

Recall  the partition function of the one matrix model:
\beqn
\label{eq:z1parti}
  Z_{N}^{(1)} \left[ \{\{ g_{\ell}\}\} \right] &=&  \int d^{N^{2}} M
e^{ - tr V\left( M; g_{\ell}\right) } \;\;\; \nonumber \\
 &=&  \int  {\displaystyle \prod_{i}^{N} } d\lambda_{i}
 \Delta\left( \lambda_{1} \cdots \lambda_{N}
\right)\Delta\left( \lambda_{1} \cdots \lambda_{N}
\right) e^{- {\displaystyle \sum_{i=1}^{N} }
 V\left( \lambda_{i} ; g_{\ell}\right) } \;\;\;, \\
 V\left(\lambda \right) &=& {\displaystyle \sum_{\ell =1}^{\ell_{max}} }
g_{\ell} \lambda_{\ell} \;\;\;.  \nonumber
\eeqn

 The
 Dyson-Schwinger equation is a set of consistency conditions on the space of
correlators which is derived under  general variations.   It can be regarded
as an equation for the space of string theories of a particular class (
in this case, the ones described by the one matrix model).
An efficient way which leads to the Virasoro constraints is the following:
 insert
$ - {\displaystyle \sum_{i=1}^{N} \lambda_{i}^{n+1} \frac{d}{d\lambda_{i}} }$
or $ {\displaystyle \sum_{i=1}^{N} \frac{1}{\zeta -\lambda_{i}} \frac{d}
{d\lambda_{i}} }$ and  express the results in terms of the couplings
and the derivatives of the couplings
in two different ways, using partial integrations. We obtain, after some
calculation  on the Vandermonde determinant,
\beqn
\label{eq:virasoro}
\hat{\ell}_{n}  Z_{N}^{(1)} &=& 0  \;\;\;,   \;\;\; n \geq -1\;\;\;,
\nonumber \\
\hat{\ell}_{n} &=& {\displaystyle \sum_{\ell=0}^{\infty}
 \ell g_{\ell} \frac{\partial}
{ \partial g_{\ell +n} } }  +
{\displaystyle \sum_{\ell=0}^{n} \frac{\partial^{2}}{\partial g_{\ell}
\partial g_{n-\ell} } } \;\;\;.
\eeqn

These are the Virasoro constraints at finite $N$. In this derivation,
it is evident that they arise from the
reparametrization  of the eigenvalue coordinates.

As a natural extension of the above derivation and also as  a warm-up
to derive  constraints of the two matrix model, consider the higher order
differential operators \cite{IM2}
\beqn
\label{eq:higherorder}
 {\displaystyle  \sum_{i} \lambda_{i}^{n}
 \left( \frac{d}{d \lambda_{i} } \right)^{m} }
  ~~~{\rm or}~~~ {\displaystyle \sum_{i} \frac{1}{\zeta - \lambda_{i}} }
\left( \frac{d}{d\lambda_{i}} \right)^{m} \;\;\;.
\eeqn

Let us introduce  the following notations
\beqn
\label{eq:notations}
j &\equiv& - {\displaystyle \sum_{\ell =1}^{\infty}
\ell g_{\ell} \zeta^{\ell-1} } \;\;\;, \nonumber \\
\frac{d}{dj} &\equiv& - {\displaystyle \sum_{\ell =0}^{\infty} \zeta^{-\ell -1}
\frac{\partial}{\partial g_{\ell}} } \;\;\;.
\eeqn

In ref. \cite{IM2},  we succeeded in writing the constraints in a closed form:
\beqn
\label{eq:higherconst}
   w^{\left( s\right)} \left( \zeta \right) Z_{N}^{\left(1\right)} &=&
 0 \;\;\;,    \nonumber \\
  s  w^{\left( s\right)} \left( \zeta \right) &=&  :  Q_{s}\left[ j(\zeta)
 + \frac{d}{dj}\left(\zeta\right) \right] :_{(-)} +
 (-)^{s}  Q_{s}^{\dagger}\left[
\frac{d}{dj}\left(\zeta\right) \right] :_{(-)} \;\;\;,
\eeqn
  where
\beq
\label{eq:Q}
   Q_{\ell} \left[ f\right] \equiv \left( \frac{\partial}{\partial \zeta}
  + f\left( \zeta\right) \right)^{\ell}
\cdot  1  \;\;\;.  ~~~~~
\eeq
The existence of such constraints is peculiar in view of the fact that
the Virasoro constraints in the double scaling limit fully characterize
a space of correlators of topological gravity \cite{VV,TOKYO}.
 We have checked \cite{IM2} in the lowest few cases
 that  these higher constraints are in fact reducible to
 the Virasoro constraints:
\beqn
\label{eq:reduce}
   &w^{\left( 2\right)} \left( \zeta \right)&  ~~~~~  {\rm Virasoro}
\;\;\; \nonumber \\
      &w^{\left( 3\right)} \left( \zeta \right)& =
 \left( j    w^{\left( 2\right)} \left( \zeta \right) \right)_{(-)}
+     \partial_{\zeta} w^{\left( 2\right)} \left( \zeta \right) \;\;\;,
\\
&\vdots&  \nonumber
\eeqn

In the approach based on the Dyson-Schwinger equation, we study a relationship
of  all possible singlet operators and  their correlation functions.
An alternative approach is the one based on  orthogonal polynomials, where
 one studies all the matrix
( both diagonal  and
off diagonal) elements
 of the  eigenvalue operator  $\hat{\lambda}$ as well as its
conjugate  $ \hat{ \frac{d}{d\lambda} } $.

Let us enumerate basic properties of the orthogonal plynomials
 $ P_{n} \left(\lambda\right) =  \lambda^{n} + c_{n-1} \lambda^{n-1}  + \cdots$
{}~~:
\beqn
\label{eq:prop}
 \lambda P_{n}\left( \lambda\right) &=& P_{n+1} +  e^{\phi_{n}} P_{n} \left(
\lambda \right) + R_{n} P_{n-1} \left( \lambda\right)  \;\;\;, \nonumber\\
\delta_{n,m} &=& \int d\lambda~ e^{-V \left(\lambda; g_{\ell} \right) }
  \frac{P_{n}\left( \lambda\right) }{ \sqrt{h_{n}} }
\frac{P_{m}\left( \lambda\right) }{ \sqrt{h_{m}} }
 \equiv  <n\mid  m>  \;\;\;, \nonumber \\
 h_{n+1} &=&  R_{n+1} h_{n}  \;\;\;, \nonumber \\
Z_{N}   &=& N! {\displaystyle \prod_{i}^{N-1} h_{i} } \;\;\;.
\eeqn
The basic equations in this approach are the matrix elements of the Heisenberg
algebra and a response of the system to a change of  parameters, namely,
  a parametric derivative of the normalization constants $h_{j}$:
\beqn
  \delta_{i,j}   &=&
 <i \mid \left[ \hat{\frac{d}{d \lambda}} , \hat{\lambda} \right] \mid
 j >  \;\;\;,   \label{eq:basics1}
 \\
 \frac{d}{dt_{k}} \ln h_{j} &=& - {\displaystyle \sum_{\ell} }
\left( \frac{\partial}{\partial t_{k}} g_{\ell} \right) <j \mid
 \hat{\lambda}^{\ell} \mid j>\;\;\;, \label{eq:basics2}
\eeqn
where  $t_{k} = t_{k} \left( g_{i} \right)$  are a set of parameters
which are functions of the original (bare) couplings $g_{i}$. The first
equation (\ref{eq:basics1})
 is regarded as a discrete version of the string equation.
The second one (\ref{eq:basics2}) may be called a discrete flow equation.
 The basic equations derived in the continuum theory are already visible
when $N$ is kept finite.

In order to demonstrate the equivalence of the two approaches, let us rederive
 the Virasoro constraints from  the equations
 (\ref{eq:basics1}),(\ref{eq:basics2}) above.
Begin with
\beqn
\label{eq:beginning}
(n+1) \hat{\lambda}^{n} = \left[ \hat{\frac{d}{d \lambda} },
\hat{\lambda}^{n+1} \right] \;\;\;.
\eeqn
   Take a trace of this relation  and multiply
by $Z_{N}$.  Namely,  $ {\displaystyle \sum_{j=0}^{N-1} <j \mid \cdots \mid
j > Z_{N}. } $  Using eq.~(\ref{eq:basics2}) $(t_{\ell} = g_{\ell})$, we find
\beqn
\label{eq:midstep}
(n+1) \frac{\partial}{\partial g_{n}} Z_{N} = {\displaystyle \sum_{\ell} }
\ell g_{\ell} \frac{\partial}{\partial g_{\ell +n}} Z_{N} -2
{\displaystyle \sum_{j=0}^{N-1}  }
 <j \mid \hat{\lambda}^{n+1} \hat{\frac{d}{d \lambda}}
\mid j>  Z_{N}  \;\;\;.
\eeqn
For the second term of the right hand side,
 use   a  formula which holds for any one-body operator
\beqn
\label{eq:lemma}
 {\displaystyle \sum_{j=0}^{N-1}} <j \mid  {\cal O} \left(  \hat{\lambda}
, \hat{ \frac{d}{d\lambda} } \right) \mid j > = <<
{\displaystyle \sum_{i=1}^{N}}
{\cal O} \left( \lambda_{i},
\frac{d}{d\lambda_{i}} \right)    >>_{\rm ave} \;\;\;,
\eeqn
where  $<<  \cdots >>_{\rm ave}$ denotes an averaging with respect to
 the partition function (eq.~(\ref{eq:z1parti})).
 ( The derivative is in between the two determinants and
 does not act on the potential.)
  Eq.~(\ref{eq:lemma})  readily follows from
\beqn
\Delta ( \lambda_{1}, \cdots \lambda_{N}) = \det \left(P_{i-1}(\lambda_{j})
\right) \;\;\;.
\eeqn
The calculation has reduced to the one mentioned  in
eq.~(\ref{eq:virasoro}).
We  obtain the Virasoro constraints  at finite $N$ again.

The properties of the orthogonal polynomials have a curious connection
\cite{Toda}
to the classical Toda lattice equation.
Let
\beqn
  \phi_{n} = \ln h_{n}   \;\;\;.
\eeqn
{}From eqs.~(\ref{eq:basics1}),(\ref{eq:basics2}),
we find a hierarchy of classical differential equations
 whose first member  is the classical Toda lattice equation:
\beqn
\label{eq:toda}
 &\left( \frac{\partial}{\partial g_{1}} \right)^{2} \phi_{n}& =
 e^{\phi_{n+1}
-\phi_{n}} -   e^{\phi_{n+1}
-\phi_{n}}   \;\;\;.
\eeqn
We leave the details to the references \cite{Toda}.
It is our hope that explicit results for the physical quantities
 come
out from this correspondence at finite $N$.

Note that  we are still dealing with  quantum theory. The feature
  that quantum correlators
 obey  classical integrable differential equations  is shared by
 other exactly solvable models of quantum field theory \cite{Korepin}.

It is instructive   that the Kazakov's original loop equation \cite{Kazakov}
 can be
directly derived from the Virasoro constraints without touching the properties
of the measure factor i.e. the Vandermonde determinant. This fact has provided
much inspiration to a recent work \cite{AIM}
 on the construction of the supersymmetric
loop
equation
 by Alvarez-Gaum\'{e}, Ma\~{n}es and the author.
Let a loop with  length $\ell$ be represented by
\beq
\label{eq:loop}
 w(\ell) =  \int \cdots  \left(
 {\displaystyle \sum_{i=1}^{N} } e^{\ell \lambda_{i}} \right)
e^{- {\displaystyle \sum_{j=1}^{N} } V\left( \lambda_{i} ;g\right) } \;\;\;,
\eeq
 where $\cdots$ denotes the measure part which we do not touch.
The operator
which Kazakov introduced \cite{Kazakov} is
\beq
\label{eq:K}
\hat{K} =  {\displaystyle \sum_{m=1}^{\infty} } m g_{m} \left
( \frac{\partial}{\partial \ell} \right)^{m-1} \;\;\;.
\eeq
The action of $\hat{K}$ on the loop turns out to be
\beqn
\label{eq:kw}
\hat{K} w(\ell) = -\frac{1}{Z_{N}} {\displaystyle \sum_{n=0}^{\infty}
\frac{\ell^{n}}{ n!} \sum_{p=0}^{\infty}  } pg_{p}
 \frac{ \partial}{
\partial g_{p+n-1} } Z_{N} \;\;\;.  \nonumber
\eeqn
Use the Virasoro constraints eq.~(\ref{eq:virasoro})
 and undo the procedure to carry out the
second derivatives of the couplings  acting on the partition function.
We find
\beqn
\label{eq:rhs}
\hat{K} w(\ell)
= << \int^{\ell}_{0} d\ell^{\prime} tr e^{\ell^{\prime} M}
 tr e^{ (\ell -\ell^{\prime}) M}  >> \;\;\;,
\eeqn
 which becomes, after using the factorization property of singlet operators
 in the planar limit,
\beq
\label{eq:loopeq}
\hat{K} w(\ell)
  = \int^{\ell}_{0} w(\ell^{\prime}) w(\ell - \ell^{\prime}) \;\;\;.
\eeq
This is Kazakov's loop equation \cite{Kazakov}.
 For   more complete discussion as well as its extension to the
 supersymmetric
loop equation and the determination of its critical points, see
the recent preprint \cite{AIM}.

Let us finally discuss how to take a double scaling limit of  the Virasoro
constraints at finite $N$ \cite{IM}
 to obtain the Virasoro constraints indexed by half
integers ( a single twisted boson) \cite{TOKYO}.
  First, the potential must be fine-tuned
in order to reach the $k$-th multicritical point of Kazakov \cite{Kazakov}
 describing
2d gravity coupled to $(2,2k-1)$ nonunitary minimal conformal matter.
Let  $a$ be the lattice spacing in the level space of the orthogonal
 polynomials.
As the leading contribution to the matrix elements of $\hat{\lambda}$ is $2$,
we rescale as
\beq
2-\hat{ \lambda} = a^{2/k} \hat{\lambda}_{sc} \;\;\;,
\eeq
with $N \rightarrow \infty$ and $a \rightarrow 0$, keeping
\beq
 g_{st} = 1/ \left(a^{2+1/k}N \right)  = {\rm finite} \;\;\;.
\eeq
In  order to drive the system slightly away from the critical point, we add
 to the original potential a source which has a nontrivial continuum limit:
\beqn
  \sum_{i} \sum_{\ell} j_{\ell} \left( 2-\lambda_{i}\right)^{\ell +1/2}
= \sum_{i} \sum_{\ell} \tilde{j}_{\ell} \lambda_{i}^{\ell} \;\;\;.
\eeqn
In the right hand side,  we reexpanded the source in terms of the polynomial
bases around the origin. This is added to the original couplings $g_{\ell}$.
We then undo the procedure to find a dressed ( renormalized ) source
expressed in terms of the original bases   $\{\{ \left( 2-\lambda_{i} \right)
^{\ell +1/2} \}\}$.    Schematically,
\beq
\label{eq:schem}
  g_{\ell}~~~ \longrightarrow~~~ \tilde{j}_{\ell} + g_{\ell}~~~ \longrightarrow
{}~~~
  j_{\ell}^{(sc)} = j_{\ell} + \cdots \;\;\;.
\eeq
We find
\beqn
\label{eq:twist}
  ~&\hat{\ell} _{n}^{(tw)}&
 Z_{N}^{(1)} \left[ \{\{ g_{\ell} \}\} + \{\{ j_{\ell}
\}\} \right] =   0 \;\;\;,~~~ n \geq 1 \;\;\;, \\
 ~&\hat{\ell} _{n}^{(tw)}& = {\displaystyle \sum_{\ell=0}^{\infty} }
\left( \ell + 1/2 \right) j_{\ell}^{(sc)} \frac{\partial}{ \partial j_{\ell +
n}^{(sc)}  }  + \frac{1}{2} {\displaystyle \sum_{\ell =1}^{n} }
\frac{ \partial^{2}}
{\partial j_{\ell-1}^{\left(sc\right)} \partial j_{n-\ell}^{\left(sc\right)}  }
\;\;\;,~~~   n \geq 1 \;\;\;, \nonumber \\
 ~&\hat{\ell} _{o}^{(tw)}& = {\displaystyle \sum_{\ell=0}^{\infty} }
\left( \ell + 1/2 \right) j_{\ell}^{(sc)} \frac{\partial}{ \partial
 j_{\ell}^{(sc)}  }  + \frac{1}{16} \;\;\;, \nonumber \\
 ~&\hat{\ell} _{-1}^{(tw)}& = {\displaystyle \sum_{\ell=1}^{\infty} }
\left( \ell + 1/2 \right) j_{\ell}^{(sc)} \frac{\partial}{ \partial j_{\ell
-1}^{(sc)}  }  +  \frac{1}{8} j_{0}^{(sc)2} \;\;\;, \nonumber
\eeqn

 This is  the Virasoro constraints of a twisted boson derived in \cite{TOKYO}
in the double scaling limit. It is worthwhile to emphasize that the limiting
process  can be postponed until the very end \cite{IM}.

\subsection {Two Matrix Model}
Let us now turn to  the two matrix model. One starts with Mehta's formula
\cite{M}:
\beqn
\label{eq:mehta}
  Z_{N}^{(2)} \left[  \{\{ g_{\ell}^{(1)} \}\}, \{\{ g_{\ell}^{(2)} \}\}
\right] &=&  \int dM_{1} dM_{2} e^{- trV^{(1)}\left( M_{1} \right) -
trV^{(2)} \left( M_{2} \right) + c trM_{1} M_{2}    }  \;\;\;, \nonumber \\
 &=& \int {\displaystyle \prod_{i} d \mu_{i} } e^{- {\displaystyle
 \sum_{i=1}^{N}  }  V^{(1)} \left( \mu_{i} \right)  } \Delta \left( \mu_{1},
\cdots  \mu_{N} \right)   \nonumber \\
   &~& \int   {\displaystyle \prod_{i} d \lambda_{i} } e^{- {\displaystyle
 \sum_{i=1}^{N}   V^{(2)} \left( \lambda_{i} \right)  + c \sum_{i}^{N} \mu_{i}
\lambda_{i} }    } \Delta \left(
 \lambda_{1},
\cdots  \lambda_{N} \right) \;\;\;.
\eeqn

This time, we first discuss the approach based on orthogonal polynomials.
Let us summarize some of the properties:
\beqn
  \delta_{ij} &=& \int d\lambda d\mu e^{- V^{(1)} \left( \mu \right)
  - V^{(2)} \left( \lambda \right)  + c\mu\lambda }
  \frac {\tilde{P}_{j} \left( \mu \right) }{\sqrt{h_{j} } }
  \frac {P_{i} \left( \lambda \right) }{\sqrt{h_{i} } }
\equiv <\tilde{j} \mid i > \;\;\;, \\
  Z_{N}^{(2)}  &=&  N! {\displaystyle \prod_{i=0}^{N-1} } h_{i} \;\;\;, \\
\lambda P_{n}  \left( \lambda \right) &=&  P_{n+1} \left( \lambda\right)
+ R_{n}  P_{n-1} \left(\lambda \right) + S_{n} P_{n-3} \left(
\lambda \right) +  \cdots  \;\;\;. \label{eq:recur}
\eeqn

Note that, unlike the case of the one matrix model, the number of terms
appearing
in the recursion relation (eq.~(\ref{eq:recur}))
 depends upon the degree of the potential.
This is a crucial observation of \cite{TY,DD}.
( See \cite{Ising,3mat}  for earlier results.)

It was Tada and Yamaguchi \cite{TY} ( and Tada \cite{Tada} later)
 who  first examined carefully the matrix elements
of the Heisenberg algebra of the two matrix model:
\beqn
\label{eq:heisen}
 \left[ \hat{\lambda} , \hat{\mu} \right] = \frac{1}{c} {\bf 1} \;\;\;.
\eeqn
They were able to determine the potentials which can produce the critical
points indexed by $ (p,p^{\prime}) = (4,5), (3,8), (3,5),$  and $(5,6)$,
precluding explicitly the earlier expectation that two matrix model contains
the critical points of the kind $(3,*)$ alone.
 The same conclusion was reached  by \cite{DD}.

Tada and Yamaguchi \cite{TY}
 explicitly took the double scaling limit of the Heisenberg algebra:
\beqn
\label{eq:dslheisen}
\hat{\mu}  &\rightarrow& X  \;\;\;,  \nonumber \\
  \hat{\lambda}  &\rightarrow& Y = X_{+}^{q/p} \;\;\;, \\
  \left[ Y, X \right] &=&  1 \;\;\;,   \nonumber
\eeqn
which is distinct from the original more heuristic analysis of
Douglas \cite{Douglas}. A prescription to obtain  $(p,p^{\prime})$ critical
points  from an asymmetric potential of degree $ 2(p^{\prime} -1)$ has been
given \cite{Tada}.
 A more general and detail
 analysis in the planar limit has been
given  recently in \cite{DEB}.

Let us now discuss   the approach based on the Dyson-Schwinger equation.
It is a useful point of view to
 regard the second set of integrations  in eq.~(\ref{eq:mehta})
 as a Laplace transform ${\cal L}$ of $ \Delta e^{-V^{(2)}}$:
\beqn
\label{eq:laplacetr}
  Z_{N}^{(2)} =  \int \prod_{i=1}^{N} e^{- {\displaystyle \sum_{j} }
 V^{(2)} \left( \mu_{j} \right)   }
  \Delta \left( \mu_{1}, \cdots \mu_{N} \right)
{\cal L} \left[ \Delta e^{-V^{(2)} } \right] \left( \mu_{1}, \cdots
\mu_{N} \right)  \;\;\;.
\eeqn

We again insert   differential operators of degree $m$
\beqn
\label{eq:diffopem}
 D_{m} \left(p\right) = \sum_{i} \frac{1}{\zeta - \mu_{i}} \left(
\frac{d}{d \mu_{i}} \right)^{m} \;\;\;.
\eeqn

Computing the action in two different ways, and using the formulas
 (eqs.~(\ref{eq:higherconst}),(\ref{eq:Q}))
developed in the one matrix  model,  we find
\beqn
 w\left( n,m\right)  Z_{N}^{\left(2\right)} &=& 0 \;\;\;, \label{eq:wconst1}
  \\
  w\left( n,m\right) &=& - \frac{(-)^{m}}{m+1}
  {\rm Res}_{\zeta =0} \left( \zeta^{n} : Q_{m+1}^{\dagger} \left[
j_{1} + \frac{d}{dj_{1}} \right] : \right)_{(-)}
    \nonumber \\
 &+& \frac{ (-)^{n}}{n+1} {\rm Res}_{\zeta=0} \left( \zeta^{m}
: Q_{n+1}^{\dagger} \left[ j_{2} + \frac{d}{dj_{2}} \right] : \right)_{(-)}
\;\;\;.   \label{eq:wconst2}
\eeqn

Let me suggest how to take a double scaling limit of the constraint
 eq.~(\ref{eq:wconst1}),(\ref{eq:wconst2}). To my knowledge,
this procedure has not been carried out explicitly yet.
Some progress has been made recently  \cite{Narain}.

First of all,  a set of couplings $g_{\ell}^{(1)}$ and $g_{\ell}^{(2)}$ must
be found which produces  the $(p,p^{\prime})$ critical point and we fine-tune
the potential to this point.
Insert a source term to the potential:
\beqn
\label{eq:sourcew}
   - \sum_{  \stackrel {i,\ell \neq 0}
 {~\pmod{p}}   } \left( c-\mu_{i}\right)^{\ell/p}
t_{\ell} -   \sum_{  \stackrel {i,\ell \neq 0}
 {~\pmod{p}}   } \left( c- \lambda_{i} \right)
^{\ell/p} t_{\ell} \;\;\;.
\eeqn
Express the constraint eq.~(\ref{eq:wconst1}) in terms of the ``renormalized''
source  in which the original couplings are absorbed.   By taking the limit,
we try to see  the reduction of our constraints to the $W_{p+1}$ constraints
of \cite{TOKYO} proven by \cite{FKN2}.

Other results from loop equations include \cite{Kostov}.

\subsection{ Discussion}
Here, we will
 discuss a few scattered items which deserve further investigations.
 In principle,  one can  apply the above method of deriving constraints
to the $c=1$ model. In practice, this is not such
an easy task: we have to introduce
 a set of time dependent couplings to the potential of the one
 dimensional matrix model.
 This would lead to the functional differential constraints on the
partition function even if we could write them down in a closed form.
If  we restrict our attention to time-independent ( static or zero momentum)
quantities alone,   it is possible to derive useful constraints
 \cite{DanGross}.  The basic quantity of interest is the density
of states
\beqn
 \rho \left( \mu \right) &=& \frac{1}{N} {\rm tr}
\delta \left( \mu - \hat{h} \right)   \;\;\; \nonumber\\
   &=& - \frac{1}{\pi N} {\rm Im} \sum_{n=0}^{\infty} <n \mid x> \left(
\frac{1}{ \mu -\epsilon_{n} +i0} \right) <x\mid n> \;\;\;.
\eeqn
rather than  the partition function.
Or consider a once integrated quantity
\beqn
   P\left(\mu\right) = -\frac{1}{\pi N} {\rm Im~tr} \ln \left( \mu -\hat{h} +
i0 \right) \;\;\;.
\eeqn
The coordinate $x$ of the one particle hamiltonian $\hat{h}$ originates
from the eigenvalues of the matrix variable.
Virasoro
 constraints are obtained from the reparametrization of eigenvalues but
 the procedure  is not
identical to that of the one matrix model \cite{DanGross}.

Our basic formulas (eqs.~(\ref{eq:higherconst}), (\ref{eq:Q}),
 (\ref{eq:wconst1}), (\ref{eq:wconst2}) ) can
 be regarded as a kind of bosonization of
 nonrelativistic fermions.  A ramification  to W geometry
 has been pursued
recently \cite{Matsuo}.

What we have been witnessing is the transmutation of the eigenvalue coordinates
into the target space degrees of freedom. This phenomenon is also present
 in  one dimensional matrix model \cite{DJ}, but conceptual understanding
seems to be still lacking.
 Constraint equations in general
 provide nontrivial  information on  target
space physics such as physics of black holes \cite{Witten}.
Clearly much more work must be dedicated on these issues together with
the issue of the definition of nonperturbative $2d$ gravity.

I thank Joe Lykken for helpful discussion.


\end{document}